\documentclass[%
 reprint,
 amsmath,amssymb,
 aps,
]{revtex4-1}

\usepackage{graphicx}
\usepackage{dcolumn}
\usepackage{bm}
\usepackage{xcolor}
\usepackage{hyperref}
\usepackage[ruled]{algorithm2e}

\begin{document}

\preprint{APS/123-QED}

\title{Real-Space Renormalization of Stabilizer Rényi Entropies in Spin Chains}

\author{Sonja Gombar}
\affiliation{Department of Physics, Faculty of Sciences, University of Novi Sad, Trg Dositeja Obradovi\'{c}a 4, 21000 Novi Sad,  Serbia}
\author{Petar Mali}
\affiliation{Department of Physics, Faculty of Sciences, University of Novi Sad, Trg Dositeja Obradovi\'{c}a 4, 21000 Novi Sad,  Serbia}
\author{Slobodan Rado\v{s}evi\'{c}}
\affiliation{Department of Physics, Faculty of Sciences, University of Novi Sad, Trg Dositeja Obradovi\'{c}a 4, 21000 Novi Sad,  Serbia}
\author{Milica Rutonjski}
\affiliation{Department of Physics, Faculty of Sciences, University of Novi Sad, Trg Dositeja Obradovi\'{c}a 4, 21000 Novi Sad,  Serbia}
\author{Milan Panti\' c}
\affiliation{Department of Physics, Faculty of Sciences, University of Novi Sad, Trg Dositeja Obradovi\'{c}a 4, 21000 Novi Sad,  Serbia}
\author{Milica Pavkov-Hrvojevi\' c}
\affiliation{Department of Physics, Faculty of Sciences, University of Novi Sad, Trg Dositeja Obradovi\'{c}a 4, 21000 Novi Sad,  Serbia}

\date{\today}

\begin{abstract}
Stabilizer states constitute an important class of quantum states that can be generated from computational-basis states using Pauli operators and Clifford gates. Although they may exhibit substantial multipartite entanglement, quantum circuits restricted to stabilizer operations can be efficiently simulated classically and therefore cannot, by themselves, provide a quantum computational advantage. Such an advantage requires non-stabilizer resources, commonly referred to as quantum magic. In this paper, we investigate the non-stabilizerness of quantum states arising in a class of spin Hamiltonians by computing their stabilizer Rényi entropies. Using real-space renormalization-group techniques, we obtain a closed-form expression valid in the low-energy, large-distance regime. We analyze how quantum magic evolves under coarse-graining and explore its behavior across different parameter regimes and quantum phases.
\end{abstract}

\maketitle


 \section{\label{sec:level1}Introduction}

Utilizing the advantages presented by quantum phenomena in information processing tasks stands as a paramount challenge in contemporary physics. To theoretically quantify such advantages, various quantum resource theories have been developed~\cite{Chitambar_Gour_2019}. Within the framework of a resource theory, one initially identifies a set of quantum states or channels lacking a specific resource of interest, considering all remaining states or channels as resourceful for task execution. A resource monotone serves to quantify this resource.  For instance, a resource theory focusing on entanglement can be formulated~\cite{Chitambar_Gour_2019,Horodecki_Horodecki_Horodecki_Horodecki_2009}, which plays a pivotal role in quantum teleportation protocols~\cite{Bennett_Brassard_Crépeau_Jozsa_Peres_Wootters_1993,Nielsen_Chuang_2010}. 

This study centers on a resource referred to as non-stabilizerness, colloquially known as 'magic'. Magic quantifies the departure of quantum states from stabilizer states, or of quantum operations from Clifford operations. Stabilizer states are those states that can be obtained from the computational basis using both Pauli operations and Clifford gate set consisting of the $S$-gate ($S=\rm{diag}[1, exp(i\pi/2)]$), Hadamard gate and CNOT gate~\cite{Haug_Kim_2023,Odavic_Haug_Torre_Hamma_Franchini_Giampaolo_2023}. By applying such circuits quantum supremacy cannot be achieved, although they can create large-scale entanglement~\cite{Harrow_Montanaro_2017}. As a resource, magic aids in establishing bounds on the classical simulability of a quantum circuit~\cite{Aaronson_Gottesman_2004, Veitch_Mousavian_Gottesman_Emerson_2014, Howard_Campbell_2017,Bravyi_Smith_Smolin_2016,difussive}. Various monotones quantifying magic have been proposed in recent years ~\cite{Howard_Campbell_2017,Heinrich_Gross_2019,Seddon_Regula_Pashayan_Ouyang_Campbell_2021}. However, their evaluation typically necessitates optimization procedures, rendering them impractical for systems beyond a few qubits. Recently, a simpler measure for computing magic, termed the stabilizer Rényi entropy (SRE), has been introduced~\cite{Leone_Oliviero_Hamma_2022PRL}. It is expressed in terms of the expectation values of Pauli strings, and significant progress has been achieved in both its numerical evaluation~\cite{Haug_Piroli_2023,huang2025fast,xiao2026exponentially,sierant2026computing} and experimental measurement~\cite{Haug_Kim_2023,Oliviero_Leone_Hamma_2022,ahmad2025experimental}. Aside from being relatively easy to calculate and experimentally measurable, the stabilizer entropies were proven to be true monotones within the context of magic-state resource theory under certain limitations \cite{Haug_Piroli_2023,Leone_Bittel_2024,esposito2026stabilizer}.

The studies of non-stabilizerness have received a wave of interest by the condensed matter communities. Note that there is a significant difference between simple magnetic insulators studied in condensed matter and qubit systems. In the latter case, exchange integrals are controllable parameters and this opens the door for their usage in construction of quantum computers \cite{Wellard,Loss}. This naturally motivates a more detailed understanding of the quantum resources hosted by many-body systems and, in particular, of the interplay between different forms of such resources. In this context, special attention has recently been devoted to the relation between entanglement and magic in many-body quantum states \cite{Liu_Winter_2022,frau2024,iannotti2026non,viscardi2025interplay,jasser2025stabilizer,turkeshi2025magic}. Furthermore, novel measures - long-range magic and quantum nonlocal non-stabilizerness - have been proposed to investigate the interplay between non-stabilizerness and quantum correlations, including entanglement \cite{Tarabunga_Tirrito_Chanda_Dalmonte_2023,esposito2026stabilizer,quantnonlnonstab,viscardi2026non}. Lately, the SRE has been used as a means to detect appearance and rise of the magic-state resources and to assess the extent to which particular spin-squeezing protocols generate non-stabilizerness \cite{2026nonstabilizerness}. The filtered stabilizer entropy has been introduced as a non-stabilizerness measure that distinguishes between  Pauli spectra of typical and atypical quantum states \cite{paulimagic}. More recently, a conformal field theoretical approach has been used to identify universal scaling properties of magic at quantum critical point \cite{magiccft}. 

At the quantum critical point a system is characterized by strong quantum fluctuations and radical changes in the ground-state properties are observed as the system goes through quantum phase transition \cite{qcp}. This makes quantum information measures invaluable tools for identifying quantum critical behavior of a model at zero temperature. Multiple quantum information measures have already been utilized as witnesses of quantum phase transitions. Namely, entanglement has been used as the most common quantum information indicator of quantum phase transition \cite{2002entanglement,Kargarian_2007,2008renormalization,Ma_Liu_Kong_2011,2025quantum}, alongside with quantum coherence \cite{2017quantum,hui2017quantum,2023renormalization} and quantum discord \cite{song2014renormalization,2016exploring}. In the present paper we employ quantum renormalization group method to investigate the role of the non-stabilizerness in detecting quantum phase transitions and evaluate the critical exponents in the cases of transverse-field Ising model and XY spin chains.

The paper is organized as follows: the model Hamiltonian and calculated quantities are introduced in Section \ref{II}, while the results are given in Section \ref{III}. Finally, Section  \ref{IV} concludes the paper.

\section{Model and Methods} \label{II}
\subsection{Stabilizer R\' enyi Entropies}

\textbf{Pure state SRE definition.} Stabilizer R\' enyi entropies (SREs)~\cite{Leone_Oliviero_Hamma_2022PRL} for pure state $\vert \psi \rangle$ are defined as 
\begin{align}
    \mathcal{M}_{\alpha} (\Psi) = \dfrac{1}{1 -\alpha} \log_{2}{ \left( \sum\limits_{\mathcal{P}_{N}} \dfrac{\vert {\rm Tr} (\Psi \mathcal{P}_{N}) \vert^{2 \alpha} }{d^{N}} \right) },\label{magicformula}
\end{align}
where $ \Psi = \vert \psi \rangle \langle \psi \vert $, $d = 2$  is the local Hilbert dimension of $N$ qubits, and $\alpha$ denotes the R\'{e}nyi index. The sum on the right-hand side runs over all possible Pauli strings built from the identity $I$ and Pauli operators $X,Y,Z$: $
\mathcal{P}_{N}=P_1\otimes P_2\otimes\cdots\otimes P_N,
\quad
P_j\in\{I,X,Y,Z\}, \quad j\in\lbrace 1,2,3,...,N\rbrace, 
$ which are referred to as phase-free Pauli strings. The Hilbert space of $N$ spins grows exponentially as $2^{N}$, while the number of Pauli strings in the Pauli group grows as $4^{N}$ and the evaluation of SREs implies a summation over an exponential number of expectation values over all possible up-to $N$ Pauli string correlation functions.  SRE can be interpreted as the R\'{e}nyi entropy of the probability distribution~\cite{Tarabunga_Tirrito_Chanda_Dalmonte_2023}
\begin{align}
    \Xi_{\mathcal{P}_{N}} (\Psi) = d^{-N} \vert {\rm Tr} (\Psi \mathcal{P}_{N}) \vert^2, \label{probb}
\end{align}
that is properly normalized
\begin{align}
    \sum\limits_{\mathcal{P}_{N}} \Xi_{\mathcal{P}_{N}}  (\Psi) = 1.
\end{align}

\textbf{Mixed state SRE definition.} For mixed states and R\'{e}nyi index $\alpha = 2$ the SRE is  defined as
    \begin{align}
        \mathcal{M}_{2} (\rho) = - \log_{2} \left( \dfrac{\sum_{\mathcal{P}_{N}} {\rm Tr}^4 (\mathcal{P}_{N}\, \rho) }{ \sum_{\mathcal{P}_{N}} {\rm Tr}^2 (\mathcal{P}_{N}\, \rho) } \right), \label{mixedSRE}
    \end{align}
where with $\rho$ we denote density matrix of the mixed state. The SREs admit the following important properties: (i) faithfulness $M_{n} (\Psi) = 0$ iff $\Psi$ is a stabilizer state, (ii) stability under Clifford unitaries $C$ as $M_{n} (C \Psi C^{\dagger}) = M_{n} (\Psi)$, and (iii) additivity $M_{n} (\Psi_{A} \otimes \Psi_{B}) = M_{n}(\Psi_{A}) + M_{n}(\Psi_{B})$. From the resource theory perspective, SRE measures the spread of a state in the basis of Pauli operators and is a valuable measure of non-stabilizerness. SRE effectively bounds  genuine measures of magic and remains amenable to efficient evaluation~\cite{Haug_Piroli_2023}. In quantum many-body systems away from criticality and in gapped regimes SRE was found to be well-approximated using single-point spin correlators~\cite{Oliviero_Leone_Hamma_2022, Odavic_Haug_Torre_Hamma_Franchini_Giampaolo_2023} due to the presence of local non-stabilizerness in these systems.  

\textbf{Long-range magic.} To quantify the long-range (non-local) non-stabilizerness, magic that is distributed over
arbitrary length scales, of a quantum state  we use the following quantity \cite{Fliss_2021}
\begin{align}
    L (\rho_{AB}) = \mathcal{M}_{2} (\rho_{AB}) - \mathcal{M}_{2} (\rho_{A}) - \mathcal{M}_{2} (\rho_{B}) \label{LRM}
\end{align}
where $A$ and $B$ are disjoint subsystems. $L(\rho_{AB})$ measures the degree to which magic cannot be removed by finite depth quantum circuits~\cite{White_Cao_Swingle_2021}. In the context of this work, we obtain the reduced density matrix state $\rho_{A}$ by performing a partial trace over a subsystem complement $A^{\rm{c}}$ as $\rho_{A} = {\rm Tr}_{A^{\rm{c}}} [ \rho] $.  

\subsection{Transverse-Field Ising Model (TFIM)}
\textbf{Model description.} First of all, we discuss the case of the one-dimensional transverse field Ising model (TFIM) with periodic boundary conditions (PBC) described by the Hamiltonian 
\begin{equation}
    H = J \sum\limits_{j =1}^{N} X_{j} X_{j+1} - h \sum\limits_{j = 1}^{N} Z_{j},  \label{HamDef}
\end{equation}
where the $X$ and $Z$ are the corresponding spin-$1/2$ Pauli matrix operators defined at the lattice site $j = 1,2, ..., N$. This model admits a global $\mathbb{Z}_{2}$ symmetry and undergoes a continuous phase transition for $J/h = 1$, which constitutes the critical point realizing an Ising conformal field theory with central charge $c = 1/2$~\cite{Sachdev_2011,Suzuki_Inoue_Chakrabarti_2013}. TFIM belongs to the integrable class of quantum many-body Hamiltonians. In particular, TFIM can be mapped to a free-fermion model using the Jordan-Wigner (JW) transformation, and explicitly diagonalized by a Fourier transform and a Bogoliubov rotation~\cite{Lieb_Schultz_Mattis_1961,Barouch_McCoy_1971,Mbeng_Russomanno_Santoro_2020}. 

\subsection{XY Model}

\textbf{Model description.} We will also focus on the following spin-1/2 periodic integrable XY chain with $N$ sites defined by the Hamiltonian
\begin{align}
    H^{\rm XY} = \dfrac{J}{4} \sum\limits_{j = 1}^{N} \left( (1 + \gamma) X_{j} X_{j+1} +  (1 - \gamma) Y_{j} Y_{j+1} \right), \label{xyham}
\end{align}
where $J$ is the spin coupling strength and $\gamma$ the spin anisotropy parameter. This model is integrable even in the presence of a transverse global magnetic field~\cite{Lieb_Schultz_Mattis_1961}. The model undergoes a continuous phase transition for $\gamma=0$, with central charge $c=1$. The mentioned model reduces to an isotropic XX model when $\gamma=0$, whereas for $\gamma=1$ it becomes the Ising model.  We limit ourselves to the case of zero transverse magnetic field, where the real-space renormalization group method is easily applicable.

\subsection{Real-Space Renormalization Group Method}\label{TFIMMM}
Real-space renormalization group (RSRG) methods play a crucial role in understanding complex systems in physics, particularly in the study of phase transitions. Originally developed for classical statistical mechanics, these techniques have been successfully extended to quantum systems, including those at zero temperature, where quantum fluctuations dominate \cite{Efrati_2014}. We will use the RSRG technique to obtain a closed-form expression for the SRE, long-range SRE, stabilizer nullity, and finally the correlation length critical exponent for the ground states of the  TFIM and XY chains. 

The core of RSRG schemes, as applied to quantum many-body Hamiltonians, focuses on generating appropriate methods for rescaling the ground and low-lying excited states. This approach has proven invaluable in studying the behavior of quantum systems near critical points; see Ref.~\cite{Suzuki_Inoue_Chakrabarti_2013} and references therein.

\section{RESULTS} \label{III}

\subsection{Transverse-Field Ising Model}

 \begin{figure}[h!]
    \centering
    \includegraphics[width=\columnwidth]{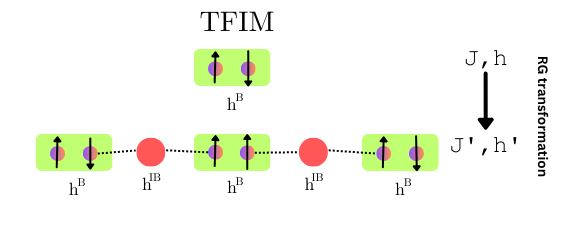}
    \caption{Schematic representation of the real-space renormalization group (RSRG) procedure for the one-dimensional transverse field Ising model (TFIM). Green rectangles represent basic blocks ($h^{\rm B}$), each containing two lattice sites (purple circles) with arrows indicating spin orientations. Red circles between blocks denote inter-block interactions ($h^{\rm IB}$). Dotted lines show connections between blocks. The right side illustrates the RG transformation, where $J$ and $h$ transform to $J'$ and $h'$, respectively. The explicit forms of the block Hamiltonian and inter-block Hamiltonian are detailed in Eq.~\eqref{form}. This RSRG scheme demonstrates the coarse-graining process used to study the system's behavior at different length scales. 
    }\label{plot0}
\end{figure}

To apply the RSRG method according to Kadanoff's blocking procedure, we write the Hamiltonian of the transverse field Ising model (TFIM) in the form:
\begin{align}
H^{\rm TFIM}=H^{\mathrm{B}}+H^{\mathrm{IB}}=\sum_{l=1}^{N/2}h_l^{\mathrm{B}}+\sum_{l=1}^{N/2}h_l^{\mathrm{IB}},
\end{align}
where $H^{\rm B}$ denotes the basic block Hamiltonian, $H^{\rm IB}$ is the inter-block Hamiltonian connecting the basic blocks, and $N$ is the total number of effective sites in the system. The subscript $l$ indexes the blocks, with $h_l^{\rm B}$ and $h_l^{\rm IB}$ representing the Hamiltonians of the $l$-th block and inter-block interactions, respectively. This is depicted in  Fig.~\ref{plot0}. This formulation allows us to systematically coarse-grain the system, revealing its behavior at different length scales. In the case of the TFIM, the explicit forms of the block and inter-block Hamiltonians are
\begin{align}
    h_l^{\mathrm{B}}=J X_{l,1}X_{l,2}-h Z_{l,1}, \quad 
    h_l^{\mathrm{IB}}=J X_{l,2}X_{l+1,1}-h Z_{l,2}. \label{form}
\end{align}  
The second subscript index references the lattice spin site within the block. Note that $h_l^{\mathrm{B}}$ is chosen in this form to obtain block Hamiltonian with the double degenerate ground state suitable for implementation of blocking procedure~\cite{Mirmasoudi_Ahadpour_2019}. The ground state of Hamiltonian $h_l^{\mathrm{B}}$ is double degenerate:

\begin{subequations}
\begin{align}
    |\psi_1^{\mathrm{GS}}\rangle&=-\frac{h+\sqrt{h^2+J^2}}{J\sqrt{1+\frac{(h+\sqrt{h^2+J^2})^2}{J^2}}}|00\rangle \notag \\ 
    &+\frac{1}{\sqrt{1+\frac{(h+\sqrt{h^2+J^2})^2}{J^2}}}|11\rangle, \label{GS1}
\end{align}
\begin{align}
    |\psi_2^{\mathrm{GS}}\rangle&=-\frac{h+\sqrt{h^2+J^2}}{J\sqrt{1+\frac{(h+\sqrt{h^2+J^2})^2}{J^2}}}|01\rangle \notag \\
    &+\frac{1}{\sqrt{1+\frac{(h+\sqrt{h^2+J^2})^2}{J^2}}}|10\rangle, \label{GS2}
\end{align}
\end{subequations}
where $|\psi_1^{\mathrm{GS}}\rangle$ and $|\psi_2^{\mathrm{GS}}\rangle$ are written using the computational basis, i.e. the eigenstates of the Pauli $Z$ operator, e.g. $\vert \! \uparrow\downarrow \rangle = \vert 01 \rangle$.

\begin{figure}[h!]
    \centering
    \includegraphics[width=\columnwidth]{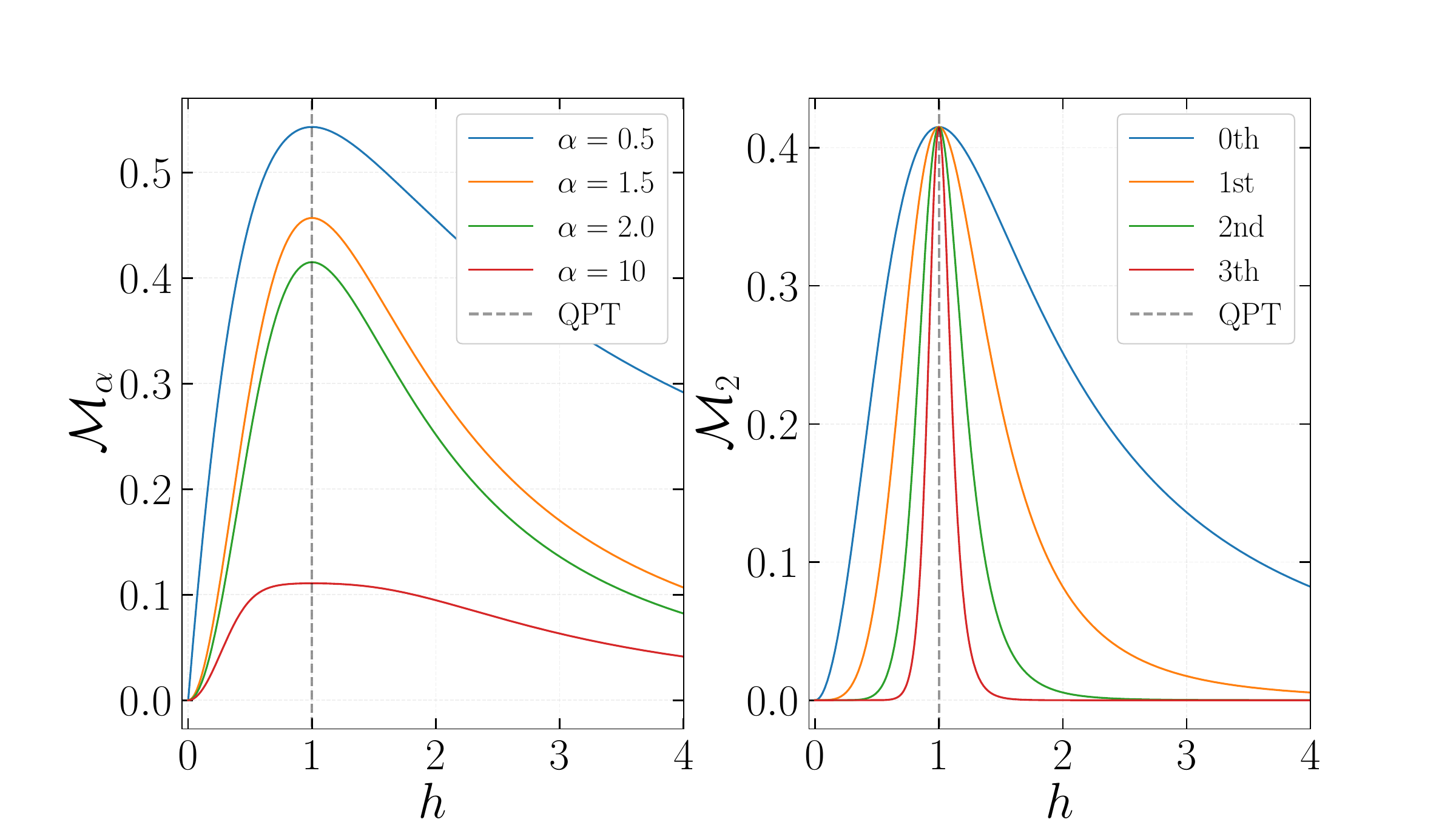}
    \caption{ \textit{Left panel}: Stabilizer R\'{e}nyi entropy (SRE) as defined in Eq.~\eqref{magicformula} of the TFIM basic block ground states from Eqs.~\eqref{GS1} -~\eqref{GS2}  with changing transverse magnetic field $h$ and R\'{e}nyi parameter $\alpha$.  \textit{Right panel}: SRE at $\alpha = 2$ for the TFIM  for different effective chain sizes (or RG steps) corresponding to the spin chain of effective length $N = 2,4,8,16$ that correspond to 0th, 1st, 2nd, 3rd RG steps, respectively. The dashed black line denotes where a continuous quantum phase transition (QPT) is located at the thermodynamic limit. We fix the coupling to $J = 1$.  
    }\label{plot3}
\end{figure}

The associated ground state energy is $E^{\mathrm{GS}}=-\sqrt{h^2+J^2}$. 
The effective Hamiltonian is constructed using the projection operator $T= \prod_{l = 1}^{N/2} T_{0}^{l}$ such that it shares the same low-lying spectrum as basic block Hamiltonian $h_{l}^{B}$. Explicitly 
\begin{align}
    T_{0}^{l} = \vert 0 \rangle_{l}  \langle \psi^{\rm GS}_{1} \vert + \vert 1 \rangle_{l}  \langle \psi^{\rm GS}_{2} \vert, 
\end{align}
where $ \vert 0 \rangle_{l}$ and $ \vert 1 \rangle_{l}$ are relabeled states of individual blocks in effective space~\cite{Martin-Delgado_Sierra_1996,Ma_Liu_Kong_2011}. The effective Hamiltonian is obtained by the transformation $H^{\rm eff} = T H^{\rm TFIM} T^{\dagger}$ from which we obtain
\begin{align}
    H^{\rm eff} =J'\sum_{l=1}^{N/2} X'_{l}X'_{l+1}-h'\sum_{l=1}^{N/2} Z'_{l},
\end{align}
where the renormalized parameters are 
\begin{align}
    J'=-\frac{J^2}{\sqrt{h^2+J^2}}, \quad
    h'=-\frac{h^2}{\sqrt{h^2+J^2}}.
\end{align}
To access the effective ground state of the model, we need to iteratively reinsert the renormalized parameters into the expressions in Eq.~\eqref{GS1} and Eq.~\eqref{GS2}. Note, however, that these effective states are not the true ground states of the spin chain. The RSRG approach only captures the low-energy limit physics corresponding to the long-distance behavior. 

Now, having access to the effective ground states we can study the behavior of SRE in this limit.  Since the basic block consists of only two spins, the SRE can be readily evaluated using Eq.~\eqref{magicformula} as

\begin{align}
\mathcal{M}_{\alpha} (J,h) &= \dfrac{1}{1 - \alpha} \log_{2}\Bigg(2^{-2\alpha}\bigg[2+\frac{2J^{2\alpha}+2h^{2\alpha}}{(h^{2}+J^{2})^{\alpha}}\bigg]\Bigg)-2. \label{TFIMSREeq}
\end{align}

Notably, this expression for the TFIM SRE exhibits non-extensive behavior; that is, the SRE of the effective ground states does not show linear growth with system size $N$, contrary to typical expectations in quantum many-body systems~\cite{Liu_Winter_2022,Odavic_Haug_Torre_Hamma_Franchini_Giampaolo_2023}. This non-extensive characteristic is anticipated, given that the fundamental block spans only two spins, and the spin coupling and magnetic field are renormalized for the effective block. In the left panel of Fig.~\ref{plot3}, we evaluate the above-obtained expression for different choices of the R\'{e}nyi parameter $\alpha$. We note in passing that SRE with $\alpha < 1$ and $\alpha > 1$ probe different aspects of non-stabilizerness present in the state~\cite{Haug_Aolita_Kim_2024}. Using the RSRG we can effectively probe arbitrary $\alpha$ parameter relevant from different points of view in terms of quantum resources.  

Using the analytic expression for the SRE we can also explicitly obtain the stabilizer nullity~\cite{Tarabunga_Tirrito_Banuls_Dalmonte_2024,Beverland_Campbell_Howard_Kliuchnikov_2020}
\begin{align}
    \nu_{0} = \lim\limits_{\alpha \to \infty} (\alpha-1) M_{\alpha} (J,h) = 1. \label{nullity}
\end{align}
Stabilizer nullity is a genuine magic monotone that characterizes the size of the stabilizer group of the considered state. For an $N$-qubit state $\vert \psi \rangle$, the stabilizer nullity is defined as
\begin{equation}
\nu_{0}(\vert \psi \rangle)
=
N-\log_{2}\left(\left|{\rm Stab}(\vert \psi \rangle)\right|\right).
\end{equation}
Here
\begin{equation}
{\rm Stab}(\vert \psi \rangle)
=
\left\{
P\in\widetilde{\mathcal{P}}_{N}:
P\vert\psi\rangle=\vert\psi\rangle
\right\}
\end{equation}
denotes the stabilizer subgroup of the full $N$-qubit Pauli group
\begin{align}
\widetilde{\mathcal{P}}_{N}
&=
\mathrm{i}^{k}P_{1}\otimes P_{2}\otimes\cdots\otimes P_{N}
, \,
k\in\{0,1,2,3\},\nonumber \\
P_j&\in\{I,X,Y,Z\}
.
\end{align}
Thus, ${\rm Stab}(\vert \psi \rangle)$ consists of those full Pauli group elements for which $\vert \psi \rangle$ is a $+1$ eigenstate. In contrast to the phase-free Pauli strings entering the definition of the SRE, the full Pauli group also includes the global phases $\pm 1$ and $\pm \rm{i}$.
In the present RSRG approach, $N$ refers to the number of qubits in the basic block and is therefore fixed to $N=2$ for the TFIM, rather than scaling with the size of the original spin chain. For generic values of the model parameters, we obtain 
$
\left|{\rm Stab}(\vert \psi \rangle)\right|=2,$ i.e. one independent nontrivial Pauli stabilizer.

For the two degenerate TFIM block ground states, the nontrivial stabilizer is the two-spin $Z$-parity operator, with its sign depending on the chosen ground state:
$Z \otimes Z\vert\psi^{GS}_{1}\rangle
=
\vert\psi^{GS}_{1}\rangle,
\qquad
Z \otimes Z\vert\psi^{GS}_{2}\rangle
=
-\vert\psi^{GS}_{2}\rangle.$ Therefore, the corresponding stabilizer groups are
${\rm Stab}(\vert\psi^{GS}_{1}\rangle)
=
\{I\otimes I,Z\otimes Z\}$,
and
${\rm Stab}(\vert\psi^{GS}_{2}\rangle)
=
\{I\otimes I,-Z\otimes Z\}$. In this way, the stabilizer nullity obtained within the RSRG approach correctly reproduces the number of independent Pauli stabilizers of the effective TFIM block ground state.

Next, we focus on the $\alpha = 2$ case, for which it was shown the SRE is a monotone for magic-state resource theory as it applies to pure states~\cite{Leone_Bittel_2024}. In this case, the expression in Eq.~\eqref{TFIMSREeq} simplifies to
\begin{align}
    \mathcal{M}_{2} (J,h) = 1 - \log_{2} {\left( 1 + \dfrac{h^4+ J^4}{(h^2 + J^2)^2}\right)}. \label{TFIMSREeq2}
\end{align}
In the right panel of Fig.~\ref{plot3}, we plot this expression in terms of the different RSRG steps providing access to larger and larger effective spin chains. Throughout the process of renormalization the basic block's SRE does not increase and peaks at the quantum critical point. This purely analytical approach shows that the SRE is maximal at the quantum critical point and with each renormalization group step (larger effective chain size) the cusp is more pronounced signaling a phase transition. Similarly to entanglement \cite{Ma_Liu_Kong_2011,Kargarian_2007,Gombar_Mali_Pantić_Pavkov-Hrvojević_Radošević_2020} the SRE and magic can resolve the critical behavior of the TFIM.  

\begin{figure}[h!]
    \centering
    \includegraphics[width=\columnwidth]{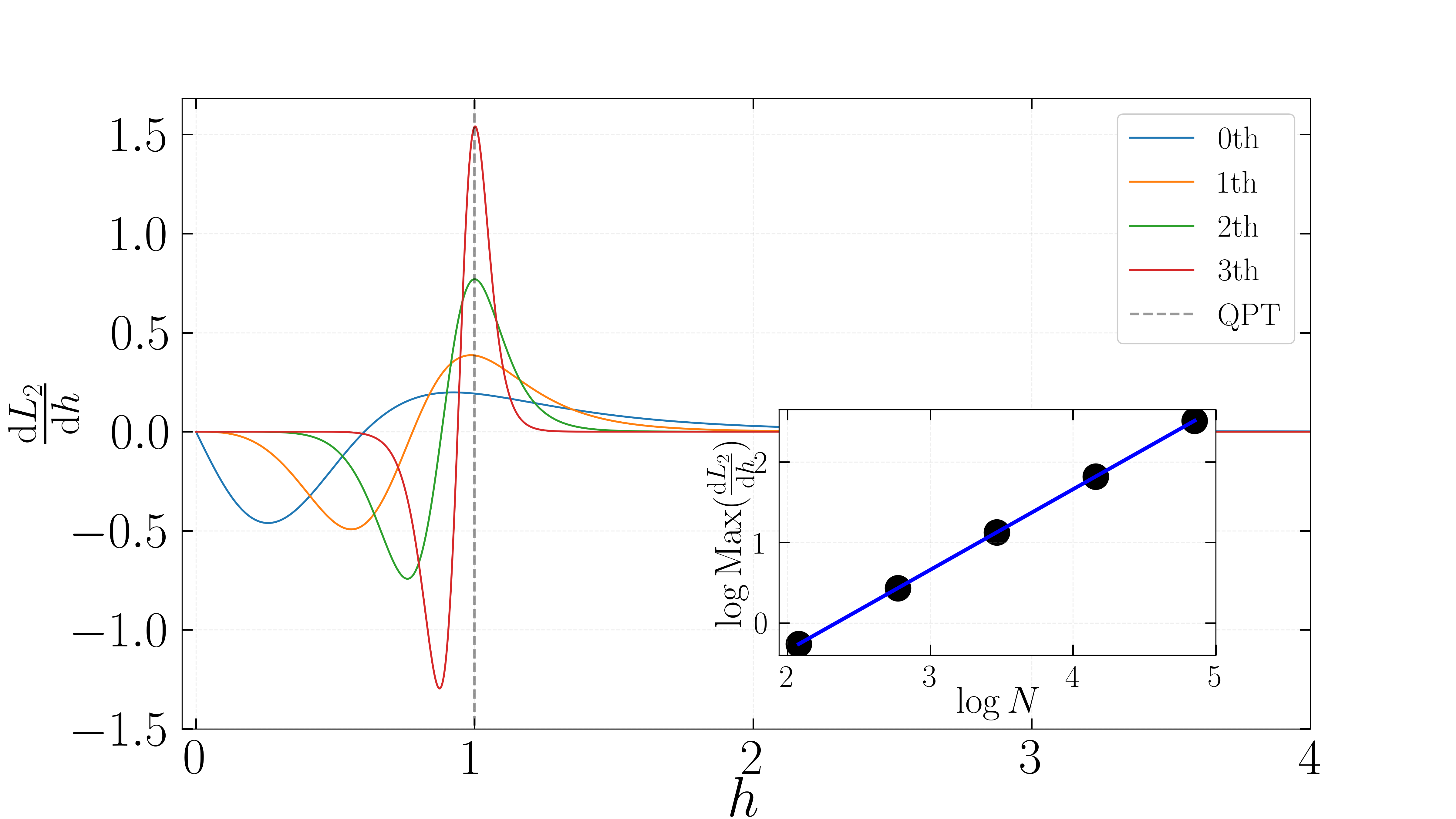}  
    \caption{ First-order derivative of the long-range SRE derived in Eq.~\eqref{long-range} with the different RG steps for the ground state of the TFIM chain. In the inset, the logarithm of the maximal value of the explicit functions presented in the main panel (i.e., the long-range magic derivative) is plotted against the increasing effective system dimension $N = 8, 16, 32, 64, 128$. With the blue line, we denote the fit to the data points with the linear function $f(x) = a + bx$, where we obtain $a = -2.34422$ and $b = 1.0014$.    
    } \label{plot4}
\end{figure}

Additionally, we obtain long-range (non-local) SRE defined in Eq.~\eqref{LRM} of the effective ground state 
\begin{align}
     L_{2}(\rho_{AB})=&-\log_2\frac{1+\left(\frac{J}{\sqrt{h^2+J^2}}\right)^4+\left(\frac{h}{\sqrt{h^2+J^2}}\right)^4}{1+\left(\frac{J}{\sqrt{h^2+J^2}}\right)^2+\left(\frac{h}{\sqrt{h^2+J^2}}\right)^2}\nonumber\\
    &+2\log_2 \frac{1+\left(\frac{h}{\sqrt{h^2+J^2}}\right)^4}{1+\left(\frac{h}{\sqrt{h^2+J^2}}\right)^2}. \label{long-range}
\end{align}
The first-order derivative of this quantity, given in Fig.~\ref{plot4}, is discontinuous at the critical point $h_{\rm c} = 1$ while the long-range SRE itself is continuous. A similar statement holds for the total pure state SRE in Eq.~\eqref{TFIMSREeq2}. This indicates the existence of a second-order phase transition in the model, to which the SRE is sensitive. Near the critical point $h_{\rm c}$, the characteristic correlation length diverges as  
\begin{equation}
\xi \sim \vert h - h_{\rm c} \vert^{-\nu}. \label{corrlength}
\end{equation} 
To extract the correlation length critical exponent $\nu$, we will first observe that after the $n$th renormalization step the correlation length near the critical point $h_{\rm c}$ exhibits the following behavior
\begin{equation}
\xi^{(n)}=\frac{\xi}{2^{n}} \sim \vert h^{(n)} - h_{\rm c} \vert^{-\nu}, \label{nthstep}
\end{equation}
where $h^{(n)}$ and $\xi^{(n)}$ are the magnetic field and the correlation length after the $n$th iteration, respectively. The number of sites per block is fixed at $2$. Taking into account the system size $N=2^{n+1}$ and Eqs. \eqref{corrlength} and \eqref{nthstep} one derives
\begin{equation}
\left|\frac{\mbox{d} h^{(n)}}{\mbox{d} h}\right|_{h_{\rm c}}\sim N^{1/\nu}.    
\end{equation} 
Accordingly,  the first-order derivative of the long-range SRE at the point corresponding to its maximal value reads
\begin{equation}
\left|\frac{\mbox{d} L_{2}}{\mbox{d} h}\right|_{h_{\rm m}}=\left|\frac{\mbox{d} L_{2}}{\mbox{d} h^{(n)}}\right|_{h_{\rm m}}\left|\frac{\mbox{d} h^{(n)}}{\mbox{d} h}\right|_{h_{\rm m}}\sim N^{1/\nu}, \label{crtitbeh}
\end{equation}
where $h_{\rm m}$ is the value of magnetic field near $h_{\rm c}$  corresponding to the maximal value of the long-range SRE. 

The inset of Fig.~\ref{plot4} shows that in the case of TFIM, the scaling behavior of the long-range SRE  first-order derivative reads
\begin{equation}
\mbox{Max} \left|\frac{\mbox{d} L_{2}}{\mbox{d} h}\right|=\left|\frac{\mbox{d} L_{2}}{\mbox{d} h}\right|_{h_{\rm m}}\sim N^{1.0014},
\end{equation}
meaning that the correlation length critical exponent $\nu=\frac{1}{1.0014}=0.9986$, according to Eq. \eqref{crtitbeh}, corresponds to the expected value for TFIM \cite{Qin_2016,Pang_2019,Kargarian_2007,Haug_Piroli_2023}. Furthermore, the SRE given in Eq. \eqref{TFIMSREeq2} exhibits analogous behavior leading to the conclusion that both SRE and long-range SRE capture the critical behavior of TFIM.

\subsection{XY Model}
\begin{figure}[b!]
    \centering
    \includegraphics[width=\columnwidth]{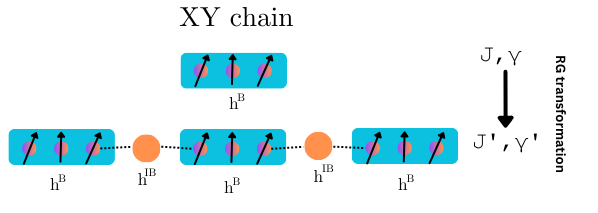}
    \caption{Schematic representation of the real-space renormalization group (RSRG) procedure for the one-dimensional XY model.  
    }\label{plot00}
\end{figure}

We now consider the spin-1/2 periodic integrable XY chain with $N$ sites with the Hamiltonian given in  Eq. \eqref{xyham}. As in the TFIM chain, our goal is to construct an effective Hamiltonian $H^{\rm eff}$ that takes into account the universal behavior of large systems (i.e. low-energy regime) using Kadanoff's block decimation approach. For this approach we define a self-similar Hamiltonian (at each application of the renormalization group procedure) by imposing a $\pi$ rotation around the $x$-axis for all even sites while leaving all the odd sites unchanged~\cite{Langari_2004}. Therefore, the transformed Hamiltonian we consider in the quantum renormalization procedure reads
\begin{align}
    H^{\rm XY}_{\rm ss} = \dfrac{J}{4} \sum\limits_{j = 1}^{N} \left( (1 + \gamma) X_{j} X_{j+1} -  (1 - \gamma) Y_{j} Y_{j+1} \right).
\end{align}
To apply the Kadanoff's approach  we write the effective Hamiltonian as
\begin{align}
    H^{\rm eff} = H^{\rm B} + H^{\rm IB} = \sum\limits_{l=1}^{N/3} h^{\rm B}_{l} + \sum\limits_{l=1}^{N/3} h^{\rm IB}_{l}, \label{eq_nec}
\end{align}
where $H^{\rm B}$ denotes the basic block Hamiltonian while $H^{\rm IB}$ is the inter-block Hamiltonian connecting the basic three-spin blocks (see Fig.~\ref{plot00}).  Explicitly, they read
\begin{align}
    h^{\rm B}_{l} &= \dfrac{J}{4}  \Big( (1+\gamma) (X_{l,1}X_{l,2} + X_{l,2} X_{l,3}) \\ \notag 
    &- (1-\gamma) (Y_{l,1} Y_{l,2} + Y_{l,2} Y_{l,3}) \Big),  \\
    h^{\rm IB}_{l} &= \dfrac{J}{4}  \left( (1 + \gamma) X_{l,3} X_{l+1,1 } - (1 - \gamma) Y_{l,3} Y_{l+1,1} \right).
\end{align}

\noindent Following the same procedure as in the case of the TFIM focusing on the low-lying spectrum we obtain the effective degenerate ground states as
\begin{align}
    \vert \Psi^{\rm GS}_{1} \rangle &= \dfrac{1}{\sqrt{2 (1 + \gamma^2)}} \Big( -\sqrt{\dfrac{1 + \gamma^2}{2}}\vert 001\rangle + \gamma \vert 010 \rangle \notag \\
    &- \sqrt{\dfrac{1 + \gamma^2}{2}} \vert 1 00 \rangle + \vert 111 \rangle \Big), \notag \\
    \vert \Psi^{\rm GS}_{2} \rangle &= \dfrac{1}{2} \Big( -\sqrt{\dfrac{2}{1 + \gamma^2}} \vert 000 \rangle + \vert 011 \rangle \notag \\
    &- \gamma \sqrt{\dfrac{2}{1 + \gamma^2}} \vert 101 \rangle + \vert 110 \rangle \Big). 
\end{align}
Under the RG procedure, the renormalized coupling and anisotropy are deduced to be of the form
\begin{align}
    J' &= \dfrac{1 + 3 \gamma^2}{2(1+\gamma^2)} J, \\
    \gamma' &= \dfrac{3 \gamma + \gamma^3}{1 + 3 \gamma^2}.
\end{align}
We explicitly obtain the SRE for the basic block of the spin-1/2 XY chain to be
\begin{align}
    \mathcal{M}_{\alpha} (\gamma) &= \dfrac{1}{1 - \alpha} \log_{2} \Bigg[  2^{-3\alpha} \bigg(2+4\left|\frac{\gamma
   ^2-1}{2(\gamma ^2+1)}\right|^{2 \alpha
   } \notag \\
   &+ 4\left|\frac{\gamma
   +1}{\sqrt{2(\gamma ^2+1)}}\right|^{2 \alpha
   }+4\left|\frac{\gamma
   -1}{\sqrt{2(\gamma ^2+1)}}\right|^{2 \alpha
   } \nonumber \\ 
   &+2\left|\frac{(1+\gamma)
   ^2}{2(\gamma ^2+1)}\right|^{2 \alpha
   }+2\left|\frac{(\gamma-1)
   ^2}{2(\gamma ^2+1)}\right|^{2 \alpha
   }\bigg)  \Bigg]-3.\label{XYSREeq}
\end{align}
By fixing $\alpha = 2$ we obtain
\begin{align}
    \mathcal{M}_{2} (\gamma) = 4 - \log_{2} {\left[  \dfrac{(3 + 10 \gamma^2 + 3 \gamma^4)^2}{(1 + \gamma^2)^4} \right]}. \label{XYSRE}
\end{align}
\begin{figure}[b!]
    \centering
    \includegraphics[width=\columnwidth]{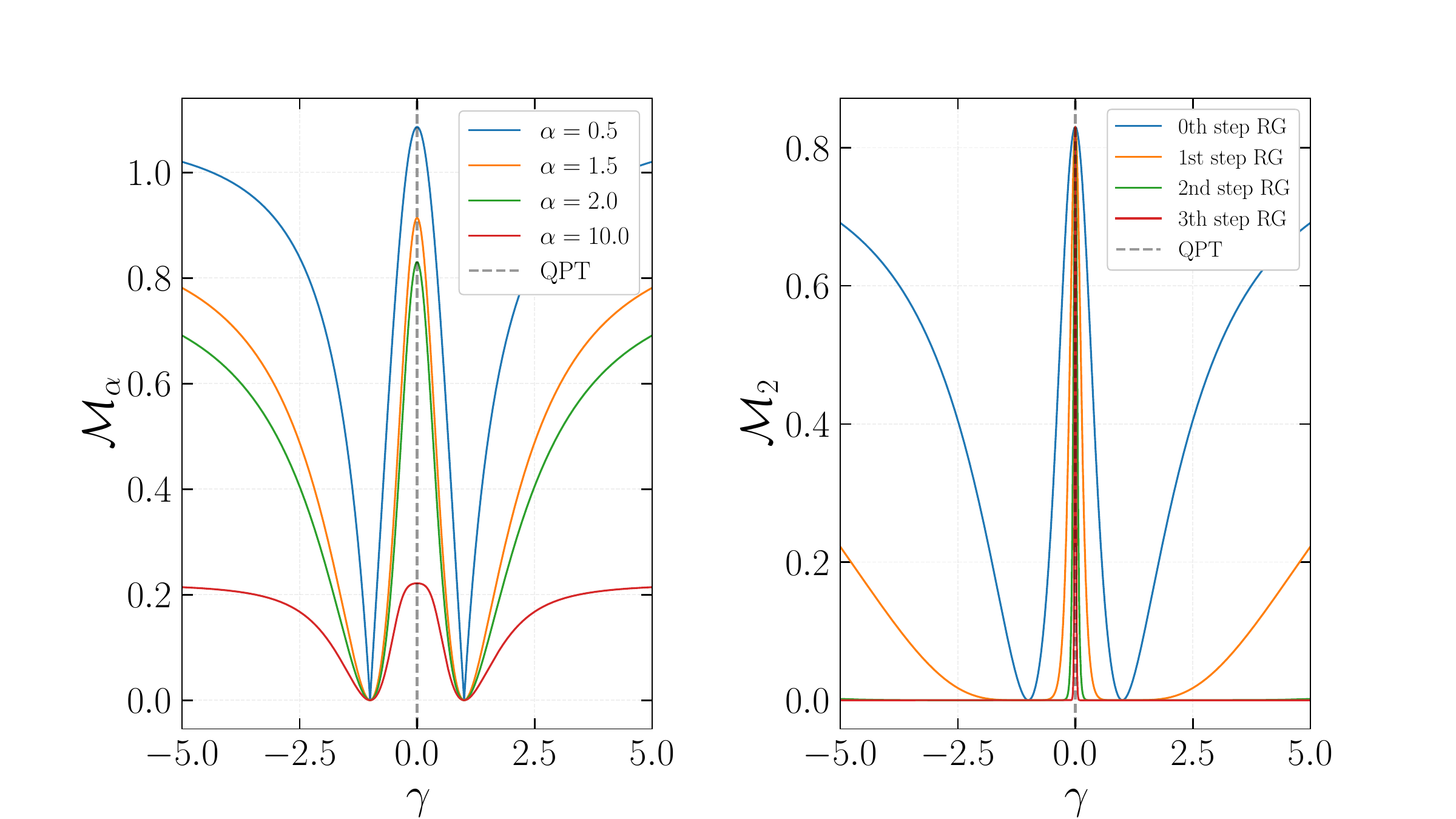}
    \caption{ \textit{Left panel}: Stabilizer R\'{e}nyi Entropy (SRE) as defined in Eq.~\eqref{XYSREeq} of the XY model basic block ground states with changing spin anisotropy parameter $\gamma$ and R\'{e}nyi parameter $\alpha$.  \textit{Right panel}: SRE at $\alpha = 2$ for the XY model  for different effective chain sizes (or RG steps) corresponding to the spin chain of effective length $N = 3,9,27,81$ that correspond to 0th, 1st, 2nd, 3rd RG steps, respectively. The dashed black line denotes where a continuous quantum phase transition (QPT) is located at the thermodynamic limit. We fix the coupling to $J = 1$.  
    }\label{plot5}
\end{figure}
These functions are presented in Fig. \ref{plot5}.

\begin{figure}[t!]
    \centering
    \includegraphics[width=\columnwidth]{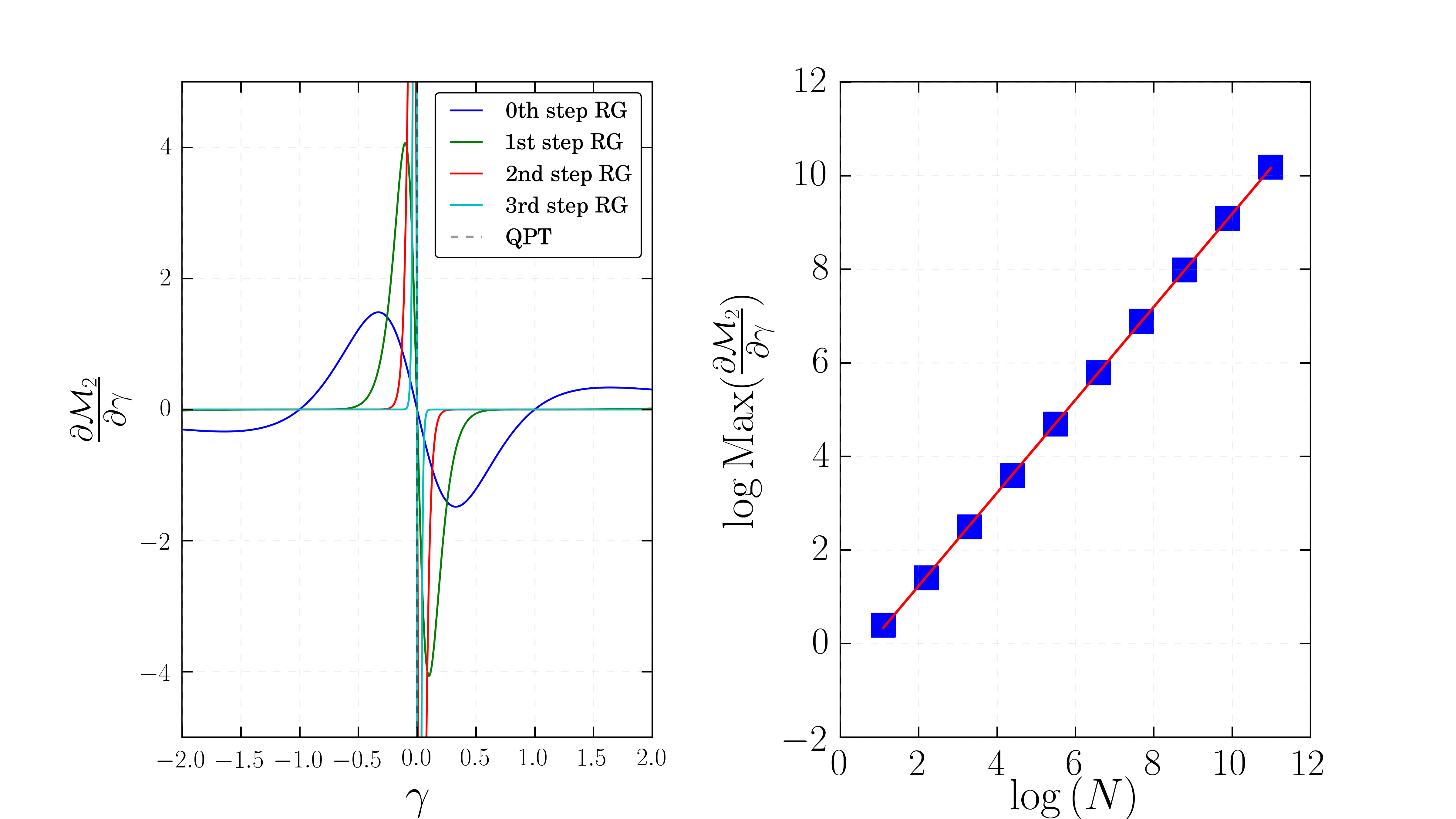}  
    \caption{\textit{Left panel}: First-order derivative of the  SRE derived in Eq.~\eqref{XYSRE} with the different RG steps for the ground state of the XY model. \textit{Right panel}: The maximal value of the explicit functions presented in the left panel against the increasing effective system dimension $N = 3, 9, 27, 81$. With the red line, we denote the fit to the data points with the linear function $f(x) = a + bx$, where we obtain $a = -0.76076$ and $b = 0.99454$.    
    } \label{plot6}
\end{figure}

Following the analogous procedure as in the case of TFIM (see Fig. \ref{plot6}), we obtain the following scaling behavior of the SRE defined by Eq. \eqref{XYSRE}
    \begin{equation}
\mbox{Max} \left|\frac{\mbox{d} \mathcal{M}_{2}}{\mbox{d} \gamma}\right|=\left|\frac{\mbox{d} \mathcal{M}_{2}}{\mbox{d} \gamma}\right|_{\gamma_{\rm m}}\sim N^{0.99454},
\end{equation}
meaning that the correlation length critical exponent is $\nu=\frac{1}{0.99454}=1.00549$ which matches the expected value for XY model \cite{He_2023}. Furthermore, we derive stabilizer nullity for the three-spin XY chain basic block from Eq. \eqref{XYSREeq} 
\begin{align}
    \nu_{0} = \lim\limits_{\alpha \to \infty} (\alpha-1) M_{\alpha} (\gamma) = 2
\end{align}
as anticipated. Note that the basic block dimension is now fixed to $N=3$. This allows us to extract the number of independent Pauli operators that stabilize the ground state of the XY model by, once again, identifying $\log_{2} {(\vert {\rm Stab} (\vert \psi \rangle )\vert )} = 1$, meaning $\vert {\rm Stab} (\vert \psi \rangle )\vert = 2$. Hence, the generic XY block ground state possesses a single independent nontrivial Pauli stabilizer. For the two degenerate three-spin ground states, the corresponding stabilizer groups are $\{I^{\otimes 3},-Z_{1}Z_{2}Z_{3}\}$ and $\{I^{\otimes 3},Z_{1}Z_{2}Z_{3}\}$, respectively, where $Z_i$ denotes the Pauli-$Z$ operator acting on the $i$th spin of the block.

\section{Conclusions and outlooks} \label{IV} 

In this paper, we make use of real-space renormalization group techniques in order to evaluate specific measures of magic and assert their connection to quantum phase transition in two specific models. Namely, we discuss the cases of one-dimensional transverse-field Ising model (TFIM) and XY chain with periodic boundary conditions. We establish that the non-stabilizer ground state can serve as a probe of quantum phase transitions, as changes in the ground-state structure across a critical point are reflected in the behavior of magic-related quantities, such as stabilizer R\' enyi entropy and long-range magic. Furthermore, we prove that the scaling behavior of the calculated quantities provides reliable estimates of the associated critical exponents.

As an additional characterization of the non-stabilizerness of the ground states, we also calculate the stabilizer nullity for both models. The stabilizer nullity provides an integer-valued characterization of the magic of a quantum state and quantifies the reduction of its stabilizer group relative to a stabilizer state. For the TFIM and XY models considered here, we obtain $\nu_0=1$ and $\nu_0=2$, respectively. This difference reflects the different block sizes rather than a difference in the number of independent stabilizer generators.

Non-stabilizerness, or magic, constitutes a fundamental resource for quantum computation beyond the stabilizer framework and is therefore of direct relevance to the pursuit of quantum advantage. Our results further demonstrate that, in quantum many-body systems, magic can play a broader role by providing a sensitive probe of quantum criticality and its universal scaling properties. This connection naturally raises the question of how magic is related to other quantum-information resources and indicators, most notably entanglement, which characterize complementary aspects of many-body quantum states. A systematic investigation of their mutual correlations, scaling properties, and possible trade-offs across quantum phase transitions therefore represents a natural extension of the present work. Such an approach could provide a more complete resource characterization of quantum critical systems and, ultimately, help identify regimes in which different quantum resources coexist or reinforce one another in ways that are advantageous for quantum information processing.

\begin{acknowledgments}
The authors gratefully acknowledge the financial support of the Ministry of Science, Technological Development and Innovation of the Republic of Serbia (Grants No. 451-03-33/2026-03/200125 \& 451-03-34/2026-03/200125). The authors would also like to thank Dr. Jovan Odavi\' c for his valuable support and insightful feedback throughout the preparation of this work. His insightful comments, constructive suggestions, and helpful discussions significantly contributed to the improvement of this manuscript.
\end{acknowledgments}

\bibliography{ref}

\end{document}